\documentclass[journal=jacsat,manuscript=letter,layout=twocolumn]{achemso}

\usepackage[T1]{fontenc} 
\usepackage{amsmath} 
\usepackage{lineno}
\usepackage{comment}
\usepackage[capitalise]{cleveref}

\author{Artem~O.~Denisov}
\email{adenisov@princeton.edu}
\author{Seong~W.~Oh}
\affiliation[Princeton Physics]
{Department of Physics, Princeton University, Princeton, NJ 08544, USA}
\author{Gordian~Fuchs}
\affiliation[Princeton Physics]
{Department of Physics, Princeton University, Princeton, NJ 08544, USA}
\author{Adam~R.~Mills}
\affiliation[Princeton Physics]
{Department of Physics, Princeton University, Princeton, NJ 08544, USA}
\author{Pengcheng~Chen}
\affiliation[IAC]
{Princeton Institute for the Science and Technology of Materials, Princeton University, Princeton, New Jersey 08544, USA}
\author{Christopher~R.~Anderson}
\affiliation[UCLA Math]
{Department of Mathematics, University of California, Los Angeles, CA 90095, USA}
\author{Mark~F.~Gyure}
\affiliation[UCLA Math]
{Center for Quantum Science and Engineering, University of California, Los Angeles, CA 90095, USA}
\author{Arthur~W.~Barnard}
\affiliation[UW]
{Department of Physics, University of Washington, 98195 Seattle, Washington, USA}
\affiliation[UW_Mat]
{Department of Materials Science and Engineering, University of Washington, 98195 Seattle,
Washington, USA}
\author{Jason~R.~Petta}
\email{petta@princeton.edu}
\affiliation[Princeton Physics]
{Department of Physics, Princeton University, Princeton, NJ 08544, USA}

\title[]{Microwave-frequency scanning gate microscopy of a Si/SiGe double quantum dot}
\abbreviations{QD,DQD,AFM,SGM}

\begin{document}


\begin{abstract}
 Conventional quantum transport methods can provide quantitative information on spin, orbital, and valley states in quantum dots, but often lack spatial resolution. Scanning tunneling microscopy, on the other hand, provides exquisite spatial resolution of the local electronic density of states, but often at the expense of speed. Working to combine the spatial resolution and energy sensitivity of scanning probe microscopy with the speed of microwave measurements, we couple a metallic probe tip to a Si/SiGe double quantum dot that is integrated with a local charge detector. We first demonstrate that a dc-biased tip can be used to change the charge occupancy of the double dot. We then apply microwave excitation through the scanning tip to drive photon-assisted tunneling transitions in the double dot. We infer the double dot energy level diagram from the frequency and detuning dependence of the photon-assisted tunneling resonance condition. These measurements allow us to resolve $\sim$65~$\mu$eV excited states, an energy scale consistent with typical valley splittings in Si/SiGe. Future extensions of this approach may allow spatial mapping of the valley splitting in Si devices, which is of fundamental importance for spin-based quantum processors.
\end{abstract}

\section{Introduction}

Quantum device performance is generally characterized in terms of a few metrics, such as the qubit relaxation time, coherence time, and gate fidelity~\cite{nielsen_chuang}. In many systems the relationship between microscopic material parameters and coherence times is poorly understood. For example, the relaxation time of superconducting qubits is limited to $\sim$100's of $\mu$s~\cite{oliver_review}. Charge noise is ubiquitous in semiconductor devices and limits the fidelity of both charge~\cite{PhysRevLett.105.246804} and spin qubits~\cite{Nichol2015}. To link microscopic materials properties and qubit performance it is desirable to develop measurement approaches that combine high spatial resolution, control of realistic quantum devices, and operation at frequencies comparable to qubit transition frequencies (typically 5 -- 20 GHz range).

Silicon spin qubits are among the leading contenders for building fault-tolerant quantum computers~\cite{PhysRevA.57.120, RevModPhys.85.961} due to their small $\sim$100$~\mathrm{nm}$ footprint~\cite{doi:10.1063/1.4922249,PhysRevApplied.6.054013} and the ability to chemically and isotopically purify the silicon host material. While long coherence times~\cite{Tyryshkin2012} and high fidelity gates~\cite{Mills_SciAdv, Xue2022, Noiri2022, HRL_2QEO} have been achieved, there are concerns about how the valley degree of freedom may impact performance as the number of silicon spin qubits scales up~\cite{PhysRevB.81.115324,2106.01391}. The strain of the Si quantum well (QW) induced by the Si and SiGe lattice mismatch partially lifts the six-fold valley degeneracy present in bulk Si~\cite{1997}. However, failure to lift the degeneracy of the low lying $\pm z$-valleys can lead to an additional uncontrolled degree of freedom~\cite{Goswami2007, PRXQuantum.2.020309, PhysRevLett.119.176803} and fast qubit relaxation~\cite{PhysRevApplied.11.044063}. An abrupt Si/SiGe interface can lift the two-fold valley degeneracy, but in reality the interface is never perfectly sharp~\cite{PhysRevB.100.125309,PhysRevB.82.205315,doi:10.1063/1.3692174,doi:10.1063/1.5033447}. Combined with atomic-scale disorder, these effects lead to a large spread in reported valley splittings 25--300 $\mu$eV~\cite{doi:10.1063/1.3569717,PhysRevLett.119.176803,PhysRevApplied.11.044063,Ferdous2018,PhysRevApplied.13.034068,PhysRevApplied.15.044033,PRXQuantum.2.020309}. Progress in better understanding the limits of valley splitting in Si/SiGe has been impeded by the lack of measurement techniques that offer spatial resolution. To accelerate improvements in heterostructure growth, it is crucial to develop a scanning probe technique capable of measuring valley splitting with spatial resolution~\cite{doi:10.1063/1.5053756}.

Scanning gate microscopy (SGM) provides spatial resolution, but it has not been applied to realistic quantum devices. In previous experiments, SGM was performed exclusively in quantum dots (QDs) defined by local anodic oxidation~\cite{PhysRevLett.93.216801,PhysRevB.83.235326} in doped GaAs or by using one layer of metallic gates~\cite{Fallahi2005}. Additionally, multiple QDs accidentally formed in carbon nanotubes~\cite{doi:10.1126/science.1069923} or semiconductor nanowires~\cite{PhysRevB.77.245327} were studied. All of these structures have an essentially open surface with at most a few thin metallic gates to deplete electrons~\cite{doi:10.1063/1.2787163}.

In this Letter we perform SGM on a device consisting of a lithographically defined double quantum dot (DQD) and an integrated QD charge sensor~\cite{doi:10.1126/science.aao5965}. The DQD is one of the most common building blocks used to define Si/SiGe charge and spin qubits~\cite{JRP_review}. To allow for the integration of the DQD in a SGM experiment, we purposely omit a gate electrode, allowing the electric potential of the SGM tip to couple to electrons in the Si QW. Using the biased tip as a movable plunger gate we demonstrate manipulation and imaging of single electrons inside the device. In the few-electron regime, where the current through the device is pinched off, we show how the charge sensor signal can be used to count electrons. By applying a microwave tone through the tip, we are able to perform excited state spectroscopy by means of photon-assisted tunneling (PAT). Our work demonstrates microscope-based characterization and control of a semiconductor quantum device and may be further extended to investigations of spin and valley coherence with spatial resolution.

\section{Results and discussion}

\begin{figure}[t]
	\includegraphics[width=1\columnwidth]{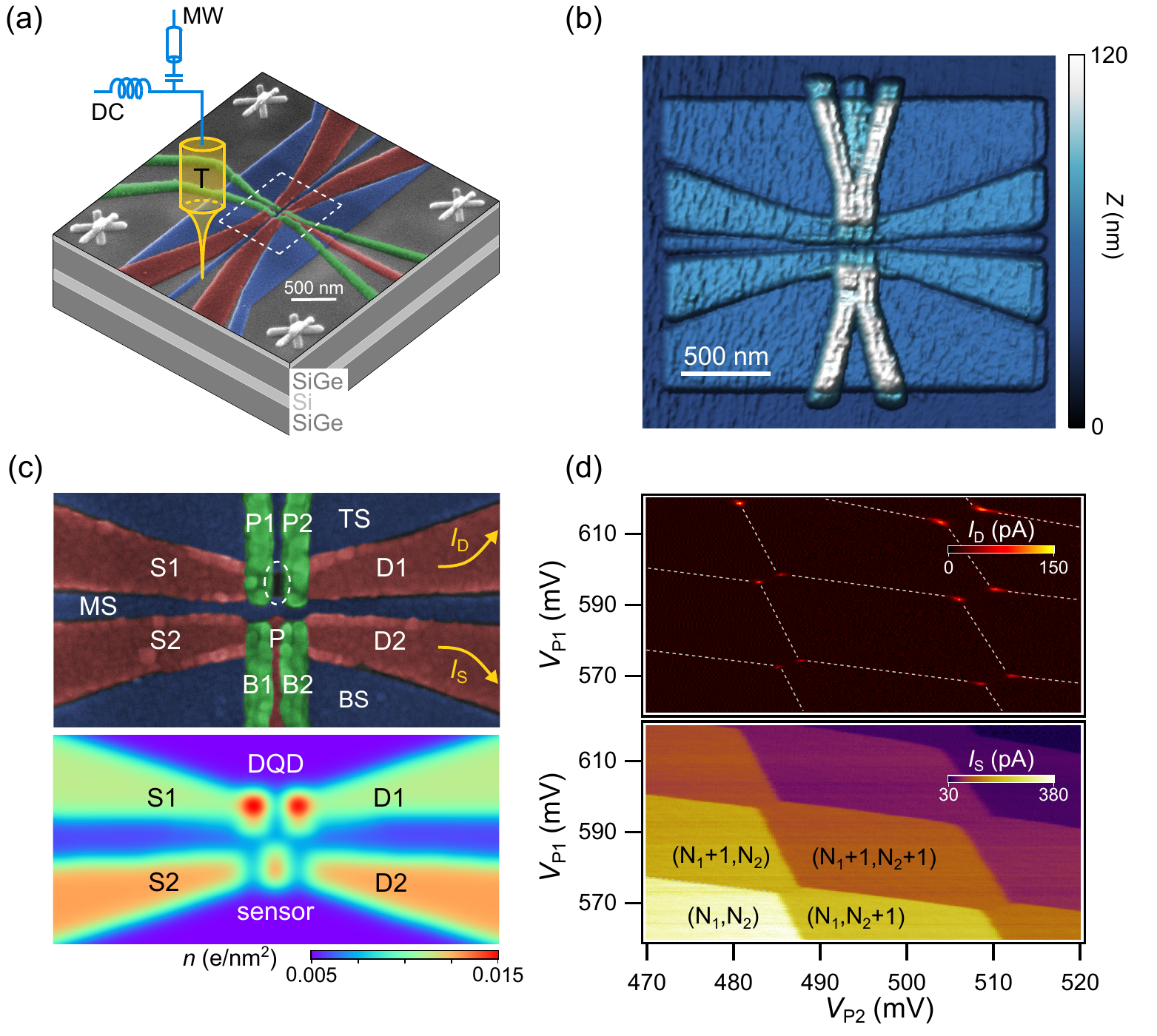}
	\caption{(a)~Schematic of the Si/SiGe device being perturbed by an atomic force microscope (AFM) tip (T). A bias-T allows microwaves to be applied to a dc-biased tip. (b)~Low-temperature AFM topography of a test structure that is adjacent to the device. (c)~(Top) False-color SEM image of the device. The absence of metal between gates P1 and P2 allows the potential of the AFM tip to perturb the electronic wave function in the Si QW. (Bottom)~Simulated charge density $n$ in the Si QW. The charge sensor is used to probe the charge occupancy of the DQD. (d)~DQD charge stability diagram extracted from the current $I_D$ flowing through the DQD~(top) and charge sensing measurements of $I_S$~(bottom).}
	\label{fig_1}
\end{figure}

Our device is fabricated on a Si/SiGe heterostructure (see Fig.~\ref{fig_1}a for a schematic of the experimental setup). The heterostructure consists of a $5~\mathrm{nm}$ thick Si QW that is buried by a $50~\mathrm{nm}$ thick layer of Si$_{\mathrm{0.7}}$Ge$_{\mathrm{0.3}}$ and capped by $2~\mathrm{nm}$ of Si. To control the electron density $n$ in the plane of the QW, we utilize a gate stack consisting of three overlapping layers of Al electrodes~\cite{PhysRevApplied.6.054013}, with progressively increasing thicknesses of 25, 45, and $75~\mathrm{nm}$. The gates define a DQD and an adjacent charge sensor QD as shown in Figs.~\ref{fig_1}b,~c. The charge sensing QD~\cite{PhysRevApplied.6.054013} is formed by means of two barrier gates (B1, B2) and a plunger gate (P). Gates S1, P1, P2, and D1 are patterned in the upper half of the device. Here we purposely omit the middle gate electrode, since its functionality is to be replaced by the SGM tip. The middle slit-gate (MS) separates the upper and lower parts of the sample, such that the DQD (sensor) current $I_{\rm D}$ ($I_{\rm S}$) can be independently measured. 

A metallic SGM tip may be used to perturb the electronic confinement potential in the Si QW. As sketched in Fig.~\ref{fig_1}a, a bias-T allows both a dc bias and microwave excitation to be applied to the tip. A typical low-temperature AFM topography image of a test structure similar to the device is shown in Fig.~\ref{fig_1}b. Here, all three layers of overlapping Al gates are visible and can be compared to the high resolution scanning electron microscope (SEM) image in Fig.~\ref{fig_1}c. All measurements were performed in cryogen-free Bluefors XLD dilution refrigerator at base electron temperature of~150~$\mathrm{mK}$. Details of the active and passive vibration damping stages utilized in the dilution refrigerator are detailed elsewhere~\cite{Seong_AIP}. We characterized three similar devices and obtained reproducible results. Data from one device are shown here.

We first discuss dc operation of the device with the SGM tip pulled far from the sample. We are able to tune the device shown in Fig.~\ref{fig_1}c to a regime where the gates P1 and P2 form a DQD. The lower panel of Fig.~\ref{fig_1}c shows the simulated charge density $n$ in the Si QW, with regions of high charge density near the ends of the S1 and D1 accumulation gates. The presence of the DQD is evident from the charge stability diagrams shown in Fig.~\ref{fig_1}d. The upper plot shows the current $I_{\mathrm{D}}$ as a function of the gate voltages $V_{\mathrm{P1}}$ and $V_{\mathrm{P2}}$ with a small source-drain bias applied across the device ($V_{\mathrm{SD}}~=~65~\mathrm{\mu V}$). As expected for weak interdot tunnel coupling~\cite{RevModPhys.75.1}, $I_{\mathrm{D}}$ is nonzero only in the vicinity of a triple point, where electrons or holes can resonantly tunnel through the DQD. The dashed lines are guides to the eye and separate different DQD charge configurations. Charge states are denoted ($N_{\mathrm{1}}$, $N_{\mathrm{2}}$), where $N_{\mathrm{1}}$ ($N_{\mathrm{2}}$) is the number of electrons in the left (right) dot. The lower panel of Fig.~\ref{fig_1}d shows the current $I_{\mathrm{S}}$ over the same range of gate voltages. As expected, $I_{\mathrm{S}}$ changes stepwise each time an electron is added to the DQD (thereby changing the total charge occupancy $N$ = $N_1$+$N_2$)~\cite{PhysRevLett.111.046801}. In contrast, the sensor is insensitive to interdot charge transitions (where $N_1$+$N_2$ is constant) due to the fact that the sensor is symmetrically placed across from the DQD~\cite{PhysRevLett.97.176803}.

\begin{figure}[t]
	\includegraphics[width=1\columnwidth]{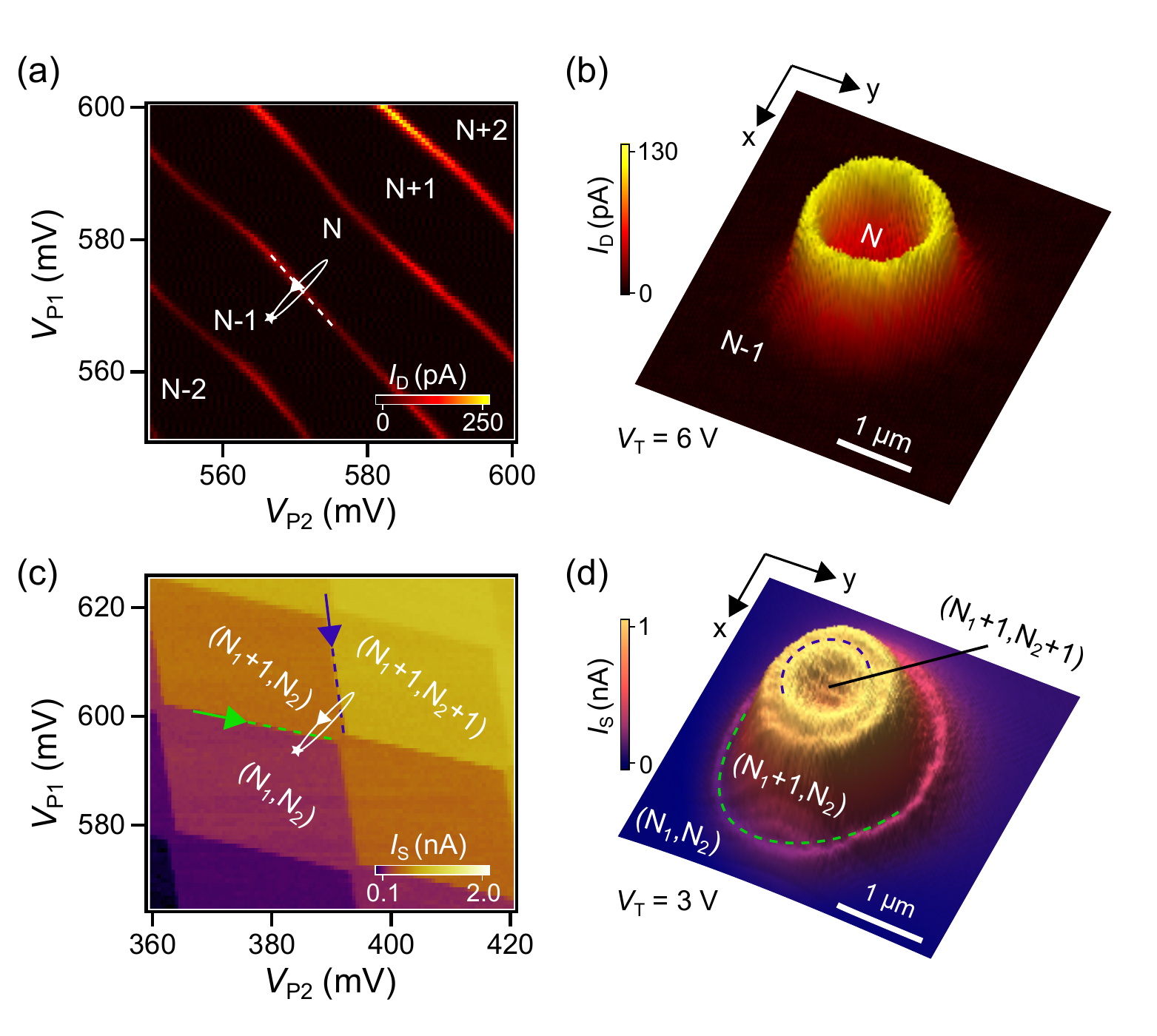}
	\caption{(a)~Charge stability diagram acquired in the single QD regime.	The star indicates the initial tuning of the device with the AFM tip pulled far away from the sample surface. The white solid line marks the conceptual trajectory of the charge state as we perform SGM. (b)~SGM image in the single dot regime at constant $V_{\mathrm{T}}~=~6~\mathrm{V}$. (c)~Stability diagram in the DQD regime. (d)~SGM image in the DQD regime at 
	constant $V_{\mathrm{T}}~=~3~\mathrm{V}$.}
	
	\label{fig_2}
\end{figure}

We now investigate the impact of the metallic SGM tip on the transport properties of the device. Large-scale scans over gate voltage space ($V_{P1}$,$V_{P2}$) show a crossover from single QD to DQD behavior~\cite{SupMat} (see SM Fig.~1). The dot current acquired in the single QD regime is plotted in Fig.~\ref{fig_2}a. Here, due to cross-capacitance effects, the device acts as a single elongated QD. To acquire SGM data, we begin with the tip far from the device and set the gate voltages in the $N-1$ electron charge state (white star in \cref{fig_2}a). The gate voltages are then held constant during SGM measurements. With the biased tip ($V_{\mathrm{T}}~=~6~\mathrm{V}$) parked $\sim$120~nm above the Al gates, we measure $I_{\mathrm{D}}$ as a function of the $xy$ coordinates of the tip, as shown in Fig.~\ref{fig_2}b. 

The electrostatic confinement potential is dependent on the ($x$,$y$,$z$) coordinates of the tip. As a result the charge stability diagram shifts to lower gate voltages (see SM Fig.~2) as we bring the positively biased tip closer to the QD. The circular feature in Fig.~\ref{fig_2}b marks the moment when the Coulomb peak separating the $N-1$ and $N$ charge states in Fig.~\ref{fig_2}a crosses the initial position in gate voltage space (marked as a white star in Fig. 2a). Equivalently, we can imagine the stability diagram unchanged and the tip instead traversing a certain trajectory in gate voltage space as shown by the white line in Fig.~\ref{fig_2}a. The almost perfectly circular feature in the SGM image defines the line of a constant tip-device interaction potential around the dot and separates the $N-1$ and $N$ charge states in the $xy$-plane~\cite{PhysRevLett.93.216801,Fallahi2005}. Sweeping $V_{\mathrm{T}}$ or moving the tip in $z$-coordinate changes the electron occupancy of the dot. These effects are well-known~\cite{doi:10.1063/1.2787163} and we discuss them in the Supplementary Material~\cite{SupMat}~. 

Operation in the few electron regime is required for the precise control of charge and spin qubits~\cite{RevModPhys.79.1217}. Now we perform the same set of measurements in the few electron DQD regime (Fig.~\ref{fig_2}c) where $I_{\mathrm{D}}$ is fully suppressed due to diminished tunneling rates and we can only rely on charge sensing. We initially park the plunger gates in the ($N_{\mathrm{1}},~N_{\mathrm{2}}$) charge state and bring the tip closer to the surface with $V_{\mathrm{T}}~=~3~\mathrm{V}$. SGM images using the sensor dot are shown in Fig.~\ref{fig_2}d. Here $I_{\rm S}$ is plotted as a function of the $xy$-position of the tip. Instead of a single Coulomb peak in Fig.~\ref{fig_2}b, here we see circular shifts in $I_{\mathrm{S}}$, each marking the boundaries between different DQD charge states as shown in Fig.~\ref{fig_2}c. The background cone-shaped signal is essentially just a Coulomb peak in the sensor dot that is electrostatically coupled to the tip. Despite the fact that we are in the DQD regime we don't observe evidence of a double-circle structure~\cite{doi:10.1126/science.1069923,PhysRevB.83.235326} in the SGM image. We attribute this result to the fact that the tip couples symmetrically to both dots. In other words, the trajectory of the initial state in the stability diagram as we move the tip radially from the center is close to a straight line in charge stability space. To quantitatively characterize the influence of the tip on the DQD we next extract the capacitance matrix.

\begin{figure}[t]
	\includegraphics[width=1\columnwidth]{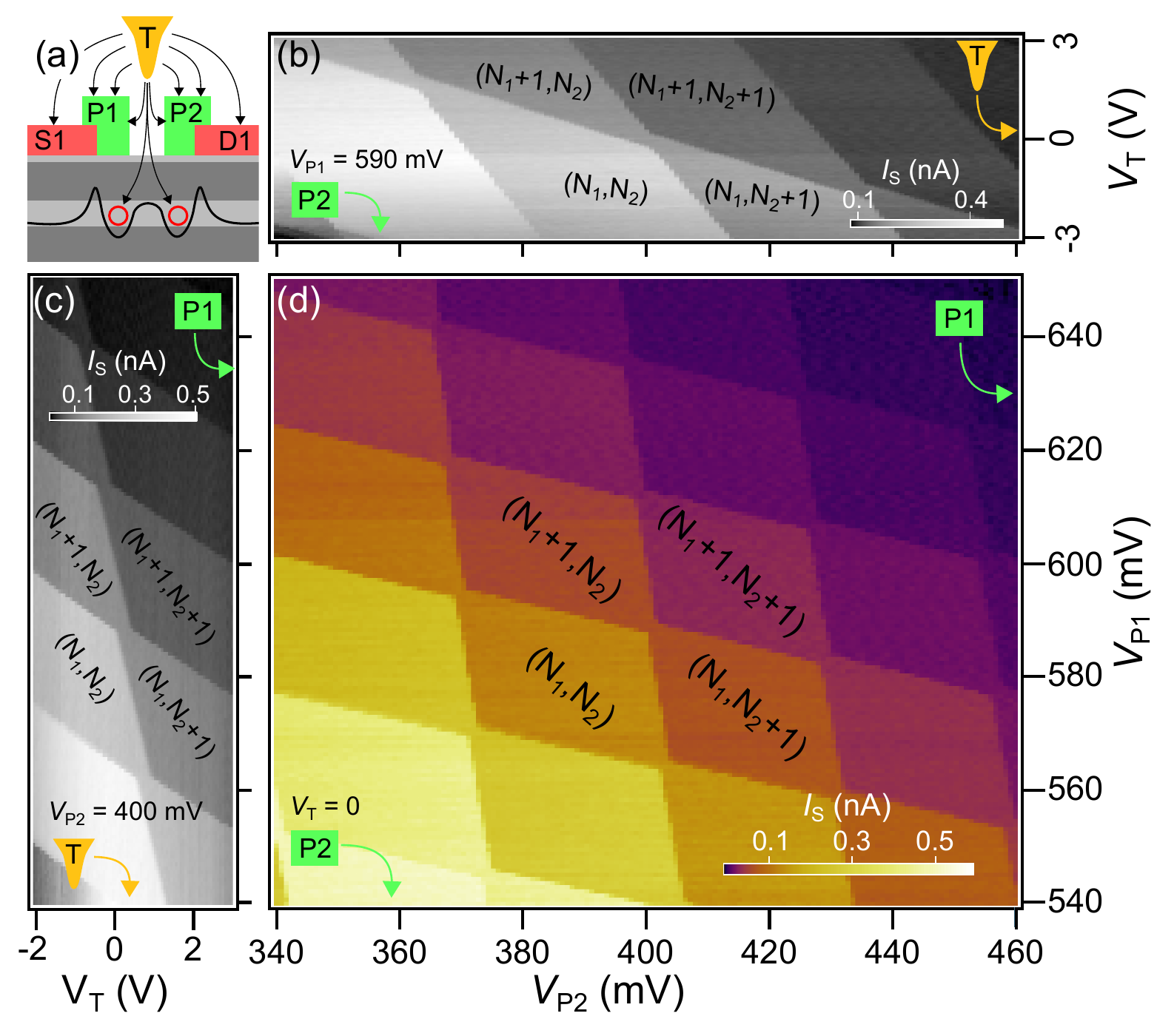}
	\caption{(a)~Illustration showing symmetric coupling between the tip and the DQD confinement potential. The metallic gates vastly screen the electric field of the tip. (b)~Stability diagram plotted as a function of $V_{\mathrm{P1}}$ and $V_{\mathrm{T}}$, with $V_{\mathrm{P2}}=400~\mathrm{mV}$. (c)~Stability diagram plotted as a function of $V_{\mathrm{T}}$ and $V_{\mathrm{P2}}$, with $V_{\mathrm{P1}}=590~\mathrm{mV}$. (d)~Stability diagram as a function of $V_{\mathrm{P1}}$ and $V_{\mathrm{P2}}$, with $V_{\mathrm{T}}=0~\mathrm{V}$.}
	\label{fig_3}
\end{figure}

As sketched in Fig.~\ref{fig_3}a, the tip is coupled almost symmetrically to both sides of the DQD. However, depending on the specific tuning of ($V_{\mathrm{P1}}$,$V_{\mathrm{P2}}$), the tip can locally serve as a plunger gate for each dot. In Figs.~\ref{fig_3}b~-~d we plot charge stability diagrams acquired by sweeping different pairs of gates, while keeping the other gate fixed. We keep the $xy$ position of the tip close to the center of the circle in Fig.~\ref{fig_2}e. All three data sets exhibit standard DQD charge stability diagrams~\cite{RevModPhys.79.1217}. From these data, we extract the capacitance matrix $\vec{C}$ from $\vec{Q} = \vec{C}\vec{V}$ (in units of $\mathrm{aF}$): 
\begin{equation}
\begin{pmatrix}
Q_{\mathrm{1}}\\
Q_{\mathrm{2}} 
\end{pmatrix}=
\begin{pmatrix}
	6.6 & 0.5 & 0.031\\
	1.3 & 5.3 & 0.023
\end{pmatrix}
\begin{pmatrix}
V_{\mathrm{P1}}\\
V_{\mathrm{P2}}\\
V_{\mathrm{T}}
\end{pmatrix}.
\end{equation}

\noindent As expected from the gate geometry, the diagonal elements in the left $2\times2$ block are a few times larger than the off-diagonal elements and a few orders of magnitude larger than tip-to-dot capacitances. The tip-to-dot capacitances are small due to the fact that the region of the 2DEG exposed to the electric field from the tip is only $80~\mathrm{nm}\times50~\mathrm{nm}$ and the rest of the electric field is screened by metallic gates as indicated in Fig.~\ref{fig_3}a.

\begin{figure}[t!]
	\includegraphics[width=1\columnwidth]{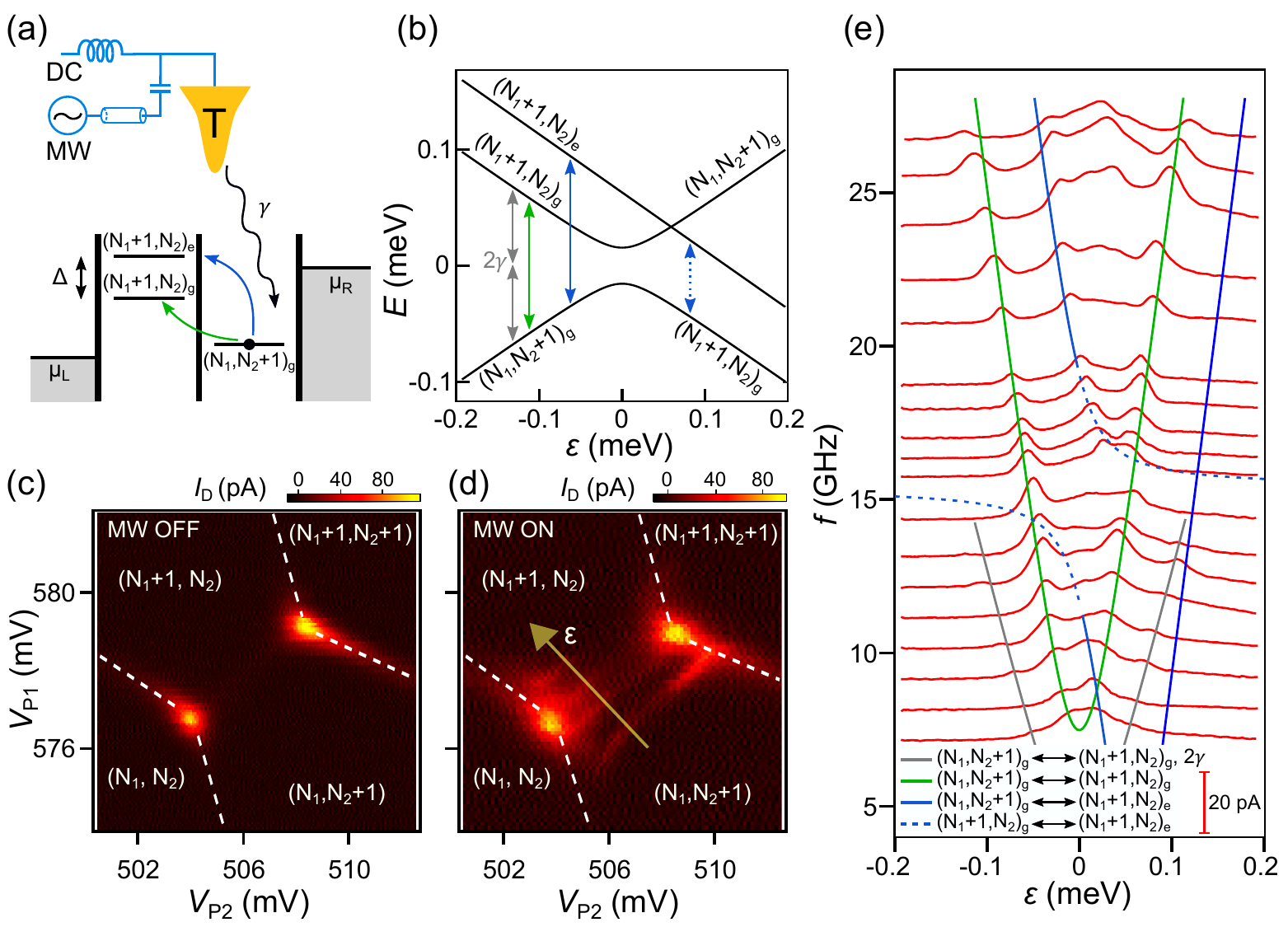}
	\caption{(a)~PAT in a DQD driven by an AFM tip. At negative detuning $\varepsilon<0$, the net current $I_{\mathrm{D}}$ through the DQD is flowing outside the triple points as the electron absorbs a photon $\gamma$ and jumps from the ground state in the left dot $(N_{\mathrm{1}},N_{\mathrm{2}}+1)_{\mathrm{g}}$ to the ground $(N_{\mathrm{1}}+1,N_{\mathrm{2}})_{\mathrm{g}}$ or excited $(N_{\mathrm{1}}+1,N_{\mathrm{2}})_{\mathrm{e}}$ state in the right dot. (b)~Three-level energy diagram of the DQD, used to fit PAT data. Two photon tunneling events are highlighted as $2\gamma$. (c,~d)~Current through the DQD around triple points with the microwave source off (c) and on ($f$ = $25~\mathrm{GHz}$) (d). Detuning axis is marked by a yellow arrow. (e)~$I_{\mathrm{D}}$ plotted as a function of $\varepsilon$ and spaced accordingly with respect to microwave frequency. Colored lines correspond to colored arrows in (b) for each transition. The data are best fit with an interdot tunnel coupling $t~=~16~\mathrm{\mu eV}$ and $\Delta~=~64~\mathrm{\mu eV}$.}
	\label{fig_4}
\end{figure}

Despite its small lever arm, we can drive PAT transitions by applying a microwave tone to the tip \cite{RevModPhys.75.1,PhysRevLett.93.186802}. PAT can be used to probe the energy level structure of the DQD, as we now demonstrate. Figure~\ref{fig_4}a illustrates the PAT process at negative detuning and with a fixed $V_{\mathrm{SD}}=\mu_{\mathrm{L}}-\mu_{\mathrm{R}}~\approx-80~\mathrm{\mu V}$. A photon ($\gamma$) or a pair of photons ($2\gamma$) incident from the tip drives the current through the DQD as the electron jumps from the ground state in the right dot to either the ground or excited state in the left dot. At positive detuning the process is similar but photon absorption is replaced by emission~\cite{Oosterkamp1998}.

The contribution of PAT processes to $I_{\mathrm{D}}$ is shown in Figs.~\ref{fig_4}c,d. With the microwave drive on (Fig.~\ref{fig_4}d) stripes of a finite current appear in the region between triple points where resonant tunneling is normally prohibited (Fig.~\ref{fig_4}c). We define the detuning axis $\varepsilon$ as being perpendicular to the PAT stripes as highlighted by a yellow arrow in Fig.~\ref{fig_4}d. The evolution of $I_{\mathrm{D}}$ linecuts is shown in Fig.~\ref{fig_4}e as a function of microwave frequency $f$. Here we plot $I_{\mathrm{D}}$, averaged over a range of microwave powers, as a function of $\varepsilon$ and spaced in accordance to microwave frequency $f$ since $I_{\mathrm{D}}\to0$ at large detuning. For $f<16~\mathrm{GHz}$, the PAT peaks are symmetric around $\varepsilon~=~0$ and shift to larger detuning with increasing photon frequency, consistent with a simple two level model of a charge qubit~\cite{PhysRevLett.93.186802}. However, for $f>16~\mathrm{GHz}$, an additional PAT peak emerges. We fit these data using a three level Hamiltonian similar to a model that includes a higher-lying valley state in one of the dots\cite{PhysRevLett.111.046801}. It includes the right dot ground state $(N_{1},N_{2}+1)_{\mathrm{g}}$, the left dot ground state $(N_{1}+1,N_{2})_{\mathrm{g}}$, and the left dot excited state $(N_{1}+1,N_{2})_{\mathrm{e}}$, as shown in Fig.~\ref{fig_4}b. The anti-crossing between ground states is governed by $hf=\sqrt{\varepsilon^{2}+4t^{2}}$, where $h$ is Planck's constant and $t$ is the interdot tunnel coupling\cite{RevModPhys.75.1}. Calculated transition frequencies are plotted in Fig.~\ref{fig_4}e along with $I_{\mathrm{D}}$. We obtain best fit values of $t~=~16~\mathrm{\mu eV}$ and an excited state energy $\Delta~=~64~\mathrm{\mu eV}$. Note that the $(N_{\mathrm{1}}+1,N_{\mathrm{2}})_{\mathrm{g}}\leftrightarrow(N_{\mathrm{1}}+1,N_{\mathrm{2}})_{\mathrm{e}}$ transition is not visible at large detuning $|\varepsilon|>2t$ since the intradot transition does not contribute to the net current. Also the $(N_{\mathrm{1}},N_{\mathrm{2}}+1)_{\mathrm{g}}\leftrightarrow(N_{\mathrm{1}}+1,N_{\mathrm{2}})_{\mathrm{e}}$ transition is not visible at large positive detuning as the population of the excited state is suppressed at low temperatures. Finally, we note that a value of $\Delta~=~64~\mathrm{\mu eV}$ is consistent with valley splittings reported in the literature\cite{doi:10.1063/1.3569717,PRXQuantum.2.020309,PhysRevLett.119.176803,Goswami2007}.

\subsection{Conclusion}
In conclusion, we demonstrated coupling between a lithographically defined DQD in a Si/SiGe heterostructure and the tip of a tuning fork based AFM. We performed manipulation and imaging of single electrons inside the DQD by means of transport and charge sensing measurements. Furthermore, we quantitatively characterized the tip-device interactions by means of the capacitance matrix, illustrating how the tip can be used as a local plunger gate to tune the electron occupancy in the DQD. By applying microwave signals to the tip, we were able to drive PAT events and perform excited state spectroscopy. Our experiments demonstrate local microscope-based dc and microwave control of a DQD - the standard building block for semiconductor quantum devices. Our work is also a starting point for spatially resolved excited state spectroscopy, which may allow us to map the variation of valley splitting in Si/SiGe heterostructures.

\begin{acknowledgement}
Supported by Army Research Office grant W911NF-15-1-0149 and the Gordon and Betty Moore Foundation’s EPiQS Initiative through Grant No. GBMF4535. The authors thank F.~F.~L.~Borjans for technical contributions to device fabrication and HRL Laboratories for providing the Si/SiGe heterostructures.
\end{acknowledgement}


\providecommand{\latin}[1]{#1}
\makeatletter
\providecommand{\doi}
  {\begingroup\let\do\@makeother\dospecials
  \catcode`\{=1 \catcode`\}=2 \doi@aux}
\providecommand{\doi@aux}[1]{\endgroup\texttt{#1}}
\makeatother
\providecommand*\mcitethebibliography{\thebibliography}
\csname @ifundefined\endcsname{endmcitethebibliography}
  {\let\endmcitethebibliography\endthebibliography}{}

\end{document}